\documentclass[prl,showpacs,twocolumn]{revtex4}

\newcommand{\expt}[1]{\left< #1\right>}
\newcommand{\Eq}[1]{Eq.~(\ref{#1})}

\newcommand{\Fig}[1]{Fig.~\ref{#1}}
\newcommand{\Figs}[1]{Figs.~\ref{#1}}
\newcommand{\Figure}[1]{Figure~\ref{#1}}
\newcommand{\Gint}{G_v^\mathrm{int}}
\newcommand{\gdot}{\dot{\gamma}}
\newcommand{\rr}{\mathbf{r}}

\newcommand{\skippa}[1]{}
\newcommand{\phiJ}{\phi_{\!J}}
\newcommand{\xhat}{\hat{x}}
\renewcommand{\v}{\mathbf{v}}
\newcommand{\Note}[1]{}

\usepackage{graphicx}

\begin{document}
\title{Diffusion and Velocity Auto-Correlation in Shearing Granular Media}

\author{Peter Olsson}

\affiliation{Department of Physics, Ume\aa\ University, 
  901 87 Ume\aa, Sweden}

\date{\today}   

\begin{abstract}
  We perform numerical simulations to examine particle diffusion at steady shear
  in a model granular material in two dimensions at the jamming density and zero
  temperature. We confirm findings by others that the diffusion constant depends
  on shear rate as $D\sim\gdot^{q_D}$ with $q_D<1$, and set out to determine a
  relation between $q_D$ and other exponents that characterize the jamming
  transition. We then examine the the velocity auto-correlation function, note
  that it is governed by two processes with different time scales, and identify
  a new fundamental exponent, $\lambda$, that characterizes an algebraic decay
  of correlations with time.
\end{abstract}


\maketitle

As the volume fraction increases in zero-temperature collections of spherical
particles with repulsive contact interaction, there is a transition from a
liquid to an amorphous solid state---the jamming transition. It has been
suggested that this transition is a critical phenomenon with universal critical
exponents \cite{Liu_Nagel} and the properties of this transition continues to be
a very active field of research.  Simulations at steady shearing have provided
strong evidence that the behavior at the jamming density actually is a critical
phenomenon \cite{Olsson_Teitel:jamming, Hatano:2008}, but questions still remain
to what extent results and ideas from ordinary critical phenomena may be taken
over to the study of jamming as well as the fundamental reason for the
observered critical behavior.

In critical phenomena the behavior is governed by a diverging length scale and
one expects that this should also be reflected in the time dependence of various
quantities.  One way to probe the time dependence is to measure the particle
displacements and thereby the diffusion constant. Experiments suggest that the
diffusion depends algebraic on the shear rate, $D\sim \gdot^{q_D}$, with $q_D<1$
\cite{Besseling_WSP, Mobius_Katgert_vanHecke}. Since this appears to be one more
critical exponent, and one usually expects relations between different critical
exponents, the existence of such a relation between $q_D$ and other exponents
that characterize the jamming transition is an interesting question.

In this Letter we examine the velocity auto-correlation function in an attempt
to understand the behavior of the diffusion constant. A careful study of this
function at very low shear rates reveals that it has both an algebraic decay and
an exponential cutoff. It futhermore turns out that these two processes are
governed by two different time scales, with the exponential decay being related
to the externally applied time scale $\sim\gdot^{-1}$ whereas the remaining
part---which we identify with an internal relaxation---is governed by a time
scale $\sim \sigma^{-1}$.

Following O'Hern \emph{et al.}\ \cite{OHern_Silbert_Liu_Nagel:2003} we simulate
frictionless soft disks in two dimensions using a Bi-dispersive mixture with
equal numbers of disks with two different radii of ratio 1.4. Length is measured
in units of the small particles ($d_s=1$). With $r_{ij}$ for the distance
between the centers of two particles and $d_{ij}$ the sum of their radii, the
interaction between overlapping particles is
\begin{displaymath}
  V(r_{ij}) = \left\{
    \begin{array}{ll}
      \frac{\epsilon}{2} (1 - r_{ij}/d_{ij})^2,\quad & r_{ij} < d_{ij},\\
      0, & r_{ij} \geq d_{ij}.
    \end{array} \right.
\end{displaymath}
We use Lees-Edwards boundary conditions \cite{Evans_Morriss} to introduce a
time-dependent shear strain $\gamma = t\gdot$. With periodic boundary conditions
on the coordinates $x_i$ and $y_i$ in an $L\times L$ system, the position of
particle $i$ in a box with strain $\gamma$ is defined as $\rr_i = (x_i+\gamma
y_i, y_i)$.  We simulate overdamped dynamics at zero temperature with the
equation of motion \cite{Durian:1995},
\begin{displaymath}
  \frac{d\rr_i}{dt} = -{C}\sum_j\frac{dV(\rr_{ij})}{d\rr_i} + y_i \gdot\; \xhat.
\end{displaymath}
The unit of time is $\tau_0 = d_s/{C}\epsilon$. We take $\epsilon=1$ and $C=1$.
We integrate the equations of motion with the Heuns method, using a time step
$\Delta t=0.2\tau_0$. As this must be considered rather large, we have checked
carefully that simulations with half that time step gives the same results to a
very high accuracy. The possibility to use such large time steps is linked to
the simple dynamics, zero temperature, and our low shear rates.

\begin{figure}
  \includegraphics[width=8cm]{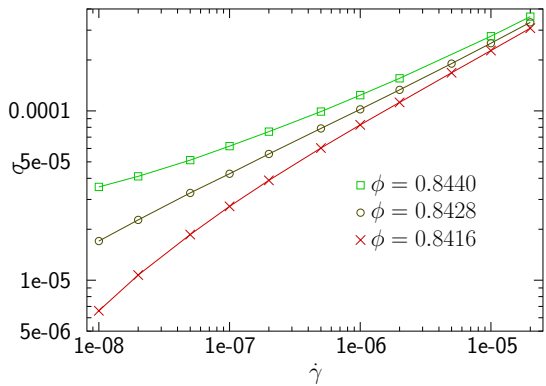}
  \caption{Behavior at $\phi_J$. At $\phi=0.8428$ which is a good candidate for
    $\phiJ$ the shear stress depends algebraically on $\gdot$ to a very good
    approximation, $\sigma\sim\gdot^{q_\sigma}$, with $q_\sigma=0.386$. Data at
    higher (squares) and lower (crosses) densities show clear curvatures.}
  \label{fig:sigma-gdot}
\end{figure}

We study a system with many particles, $N=65536$, at $\phiJ$, since the
correlation length in the system should only depend on the finite shear rate and
one therefore expects a simpler behavior. We have checked that our results are
not affected by finite size effects. The behavior of the shear stress at three
densities at and around $\phi=0.8428$ is shown in \Fig{fig:sigma-gdot}. At
$\phi=0.8428$, which is our candidate for $\phiJ$, the shear stress is algebraic
in the shear rate, $\sigma\sim \gdot^{q_\sigma}$ with $q_\sigma=0.386$, whereas
the data away from $\phiJ$ have clear curvatures. In the notation of
Ref.~\cite{Olsson_Teitel:jamming}, $q_\sigma= \Delta/(\beta+\Delta)$. We remark
that the fit is not entirely perfect, in spite of the nice algebraic behavior in
\Fig{fig:sigma-gdot}. This is the reason why the present estimate $\phiJ=0.8428$
is somewhat higher than the estimate in Ref.~\cite{Olsson_Teitel:jamming}. Our
new data (which extends down to lower shear rates) also show that a
high-precision determination of $\phiJ$ and the related exponents is a difficult
task. This is due to some corrections to the expected scaling behavior, as will
be discussed elsewhere. For the purpose of the present Letter the approximate
value $\phiJ\approx 0.8428$ is, however, entirely sufficient.

We determine the diffusion constant from the transverse displacements, i.e.\ the
displacements in the $y$ direction, and the velocity auto-correlation function
from the $y$ component of the velocity,
\begin{displaymath}
  g_v(t) = \expt{v_y(t') v_y(t'+t)},
\end{displaymath}
where the average is over all particles and a large number of initial times,
$t'$. Here and in the following, $t$ is the difference between two absolute
times.  The velocity auto-correlation function has been examined before
\cite{Ono_Tewari_Langer_Liu}, but the present data with higher precision at
lower shear rates makes it possible to do a more thorough analysis of its
properties.  The relation to the diffusion constant is given by the fundamental
relation
\begin{equation}
  D = \int_{-\infty}^\infty dt\; g_v(t) = g_v(0) \int_{-\infty}^\infty dt\;
  G_v(t),
  \label{eq:D-gv}
\end{equation}
where we introduce the normalized $G_v(t) = g_v(t)/g_v(0)$. It is convenient to
write the expression in terms of $G_v(t)$ both since it is the quantity that
will be examined below and since the prefactor, $g_v(0)$, has a known behavior,
$g_v(0) \equiv v_y^2 \sim \sigma\gdot \sim \gdot^{1+q_\sigma}$, which follows
from $N\expt{\v^2}/{C} = L^2 \sigma\gdot$ \cite{Ono_Tewari_Langer_Liu}.

\begin{figure}
  \includegraphics[width=8cm]{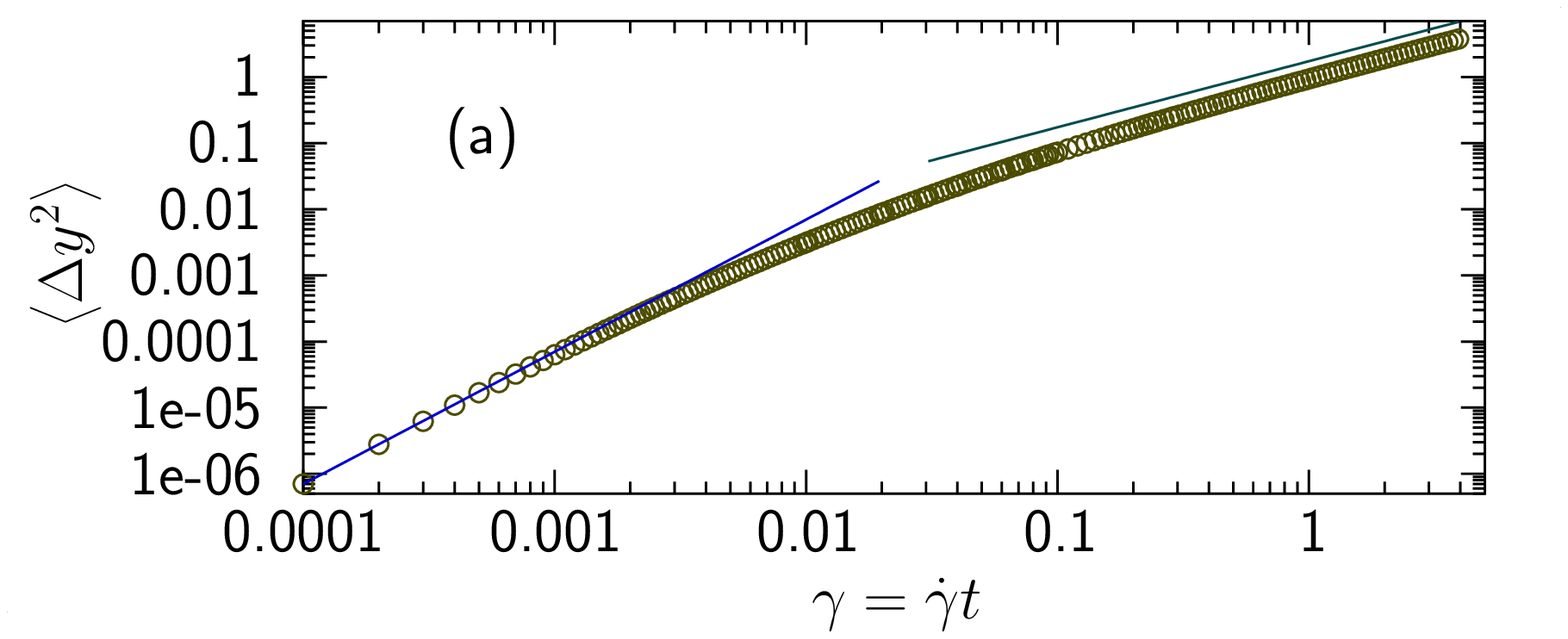} 
  \includegraphics[width=8cm]{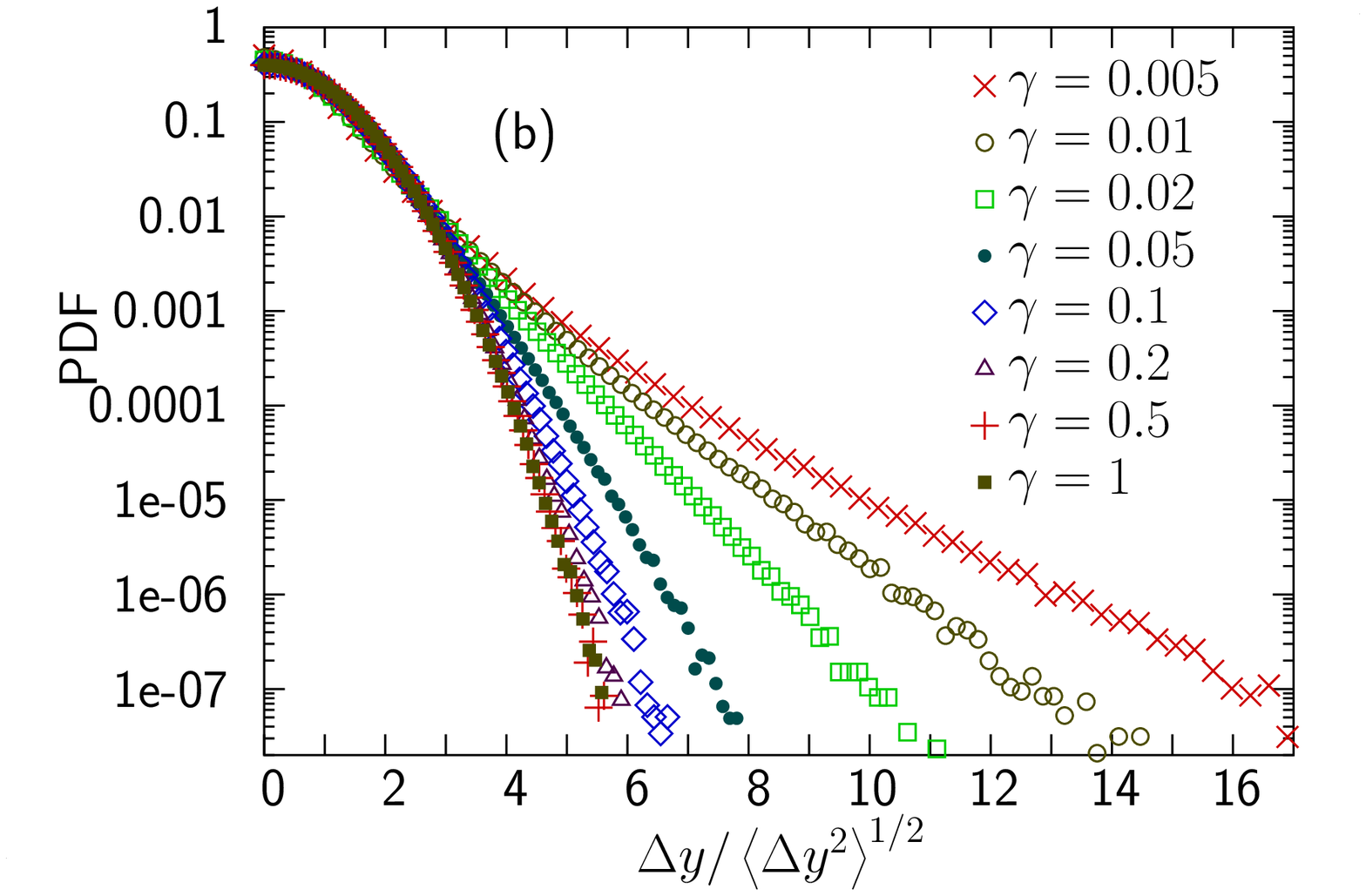} 
  \includegraphics[width=8cm]{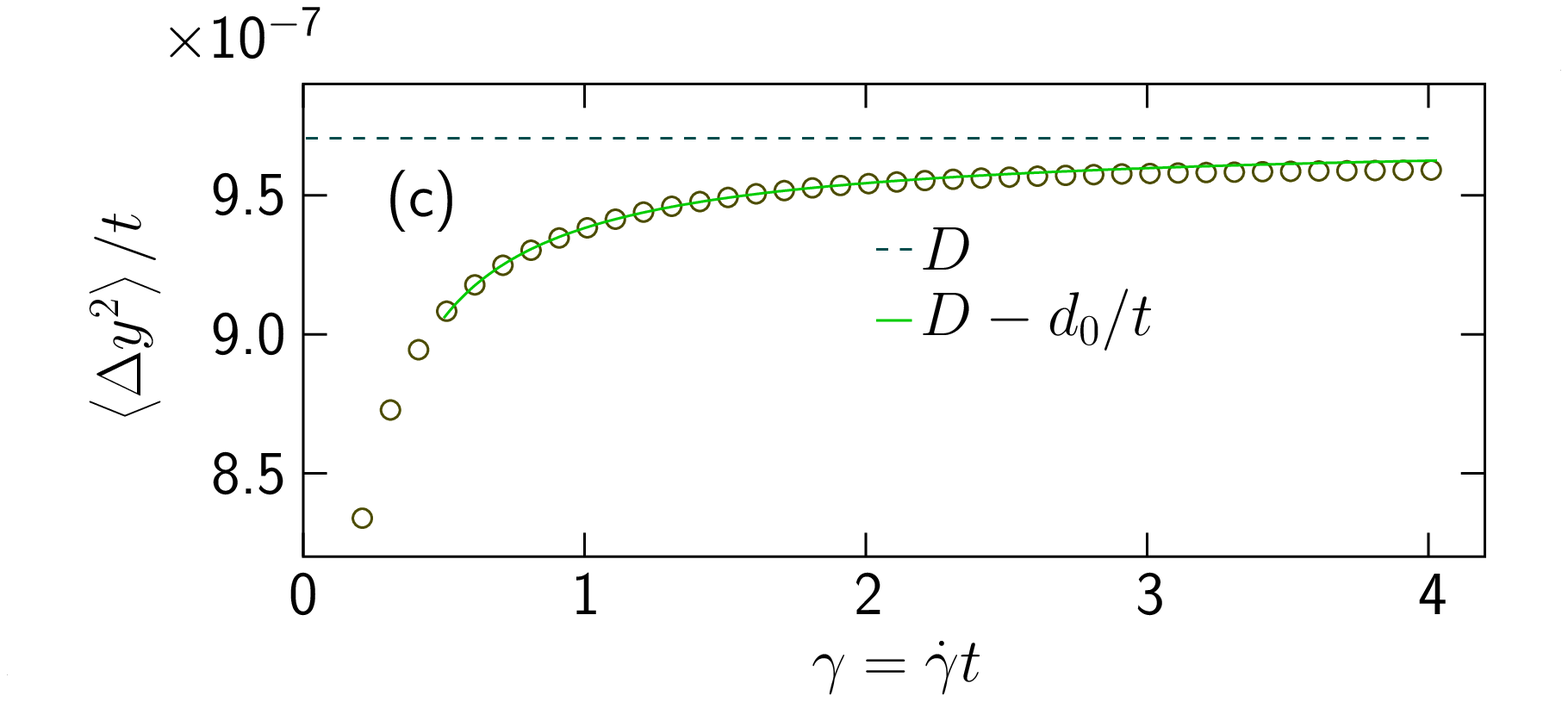} 
  \caption{Particle displacements for $\gdot=10^{-6}$. Panel (a) shows the
    crossing over of $\expt{\Delta y^2}$ from ballistic behavior at short times
    to diffusion at large times. The slopes of the solid lines are 2 and 1,
    respectively. Panel (b) shows
    the probability distribution function of particle displacements, normalized
    by the width of the respective distributions. Note the exponential shape at
    small strains (= short times) that crosses over to a Gaussian distribution
    at $\gamma\approx 0.2$. Panel (c) shows the determination of $D$ from the large
    $\gamma$ part of the same data. The solid line is from fitting $\expt{\Delta
      y^2}$ to \Eq{eq:Dy2}; the dashed line corresponds to $D$. }
  \label{fig:y}
\end{figure}

Some quantities related to the particle displacements are shown in
\Figs{fig:y}. To make it easier to interpret the figures these quantities are
plotted against $\gamma$ (the strain increment), though we discuss the behavior
in terms of $t$.  Panel (a) which shows $\expt{\Delta y^2}$ against $\gamma$,
illustrates the crossover from ballistic motion at short times to diffusion,
$\expt{\Delta y^2} \sim t$. The probability distribution function (PDF) of
$\Delta y$ (normalized by the width of the distribution), for several different
strain increments, is shown in panel (b). The PDF crosses over from exponential
behavior at short times (small $\gamma$) to a Gaussian at longer times, as found
by others \cite{Chaudhuri_Berthier_Kob,Mobius_Katgert_vanHecke}.  Our
determination of the diffusion constant is illustrated in \Fig{fig:y}(c). As the
figure shows it is difficult to determine $D$ from the long time limit of
$\expt{\Delta y^2}/t$ since this quantity approaches the constant value $=D$
very slowly. The reason for this is a remainder of the short time behavior. For
$t> t_0$, where $t_0$ is the range of the velocity correlations (such that
$G_v(t)$ may be neglected for $t\geq t_0$; we choose $\gamma_0=0.5$,
$t_0=\gamma_0/\gdot$), it is easy to show that the expression for the mean
square distance is
\begin{equation}
  \expt{\Delta y^2(t)} = \int_0^t dt' \int_0^t dt'' g_v(t'-t'')
  = Dt - d_0
  \label{eq:Dy2}
\end{equation}
with $D$ from \Eq{eq:D-gv} and $d_0 = \int_0^{t_0} dt'\int_{t'}^{t_0} dt''
g_v(t'')$. The solid line in \Fig{fig:y}(b) is from a fit to \Eq{eq:Dy2} with
data from the interval $\gamma_0 = 0.5\leq \gamma\leq2$. The dashed line is the
estimated value of $D$.

\begin{figure}
  \includegraphics[width=8cm]{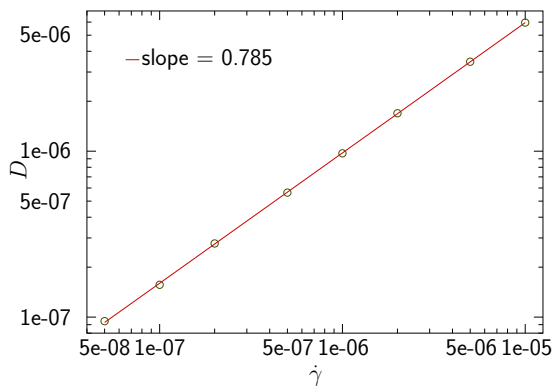} 
  \caption{Diffusion constant versus shear rate. The open circles are the
    diffusion constant at $\phi=0.8428$ versus $\gdot$. The data is well fitted
    to an algebraic relation $D\sim\gdot^{q_D}$, with $q_D=0.785(5)$.}
  \label{fig:D-gdot}
\end{figure}

\Figure{fig:D-gdot} shows diffusion constant versus shear rate, determined with
the same kind of fits. The behavior is $D\sim\gdot^{q_D}$, with $q_D=0.785(5)$.
This implies that the distance moved per unit strain \emph{decreases} with
increasing shear rate \cite{Malandro_Lacks:1998}.  The corresponding exponents
from experiments are $q_D=0.80\pm0.01$ from three dimensional colloids
\cite{Besseling_WSP} and $q_D=0.66\pm0.05$ from bubble rafts
\cite{Mobius_Katgert_vanHecke}. We note that the experiments on the colloids
were performed at a density close to the jamming density whereas the bubble raft
was studied well above $\phiJ$. This is a possible reason why the value of $q_D$
in the colloids agrees well with our value obtained at $\phiJ$.

\begin{figure}
  \includegraphics[width=4cm,bb=50 321 334 535]{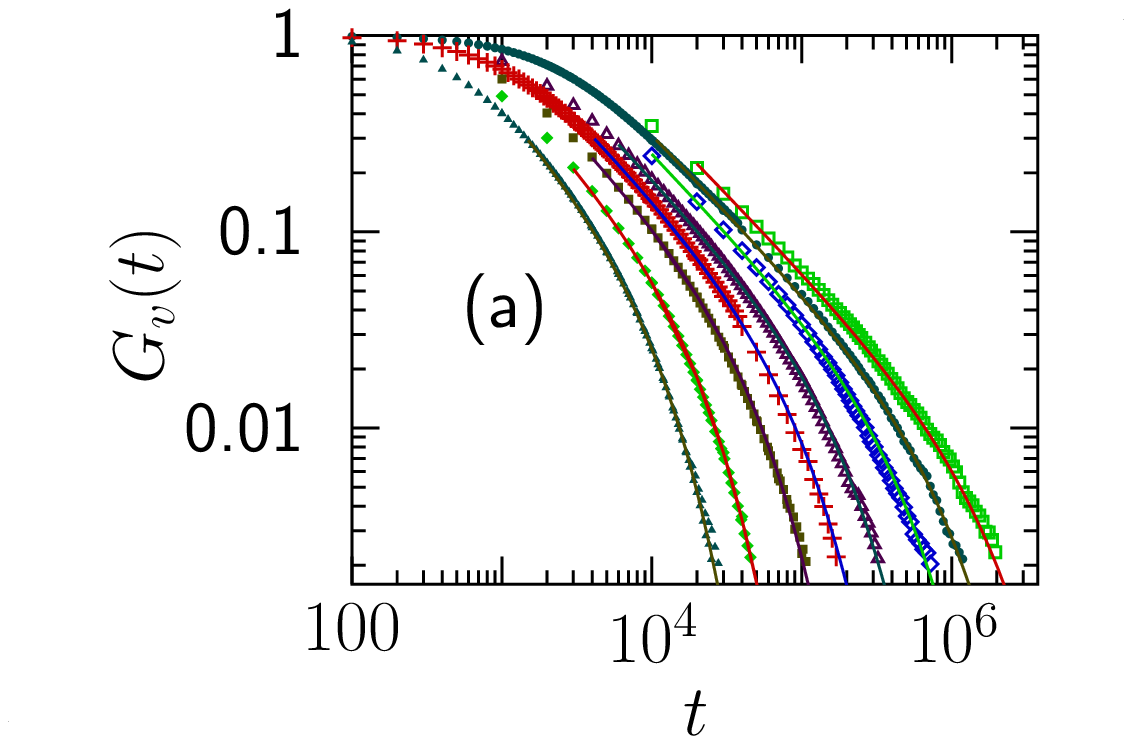} 
  \includegraphics[width=4cm,bb=50 321 334 535]{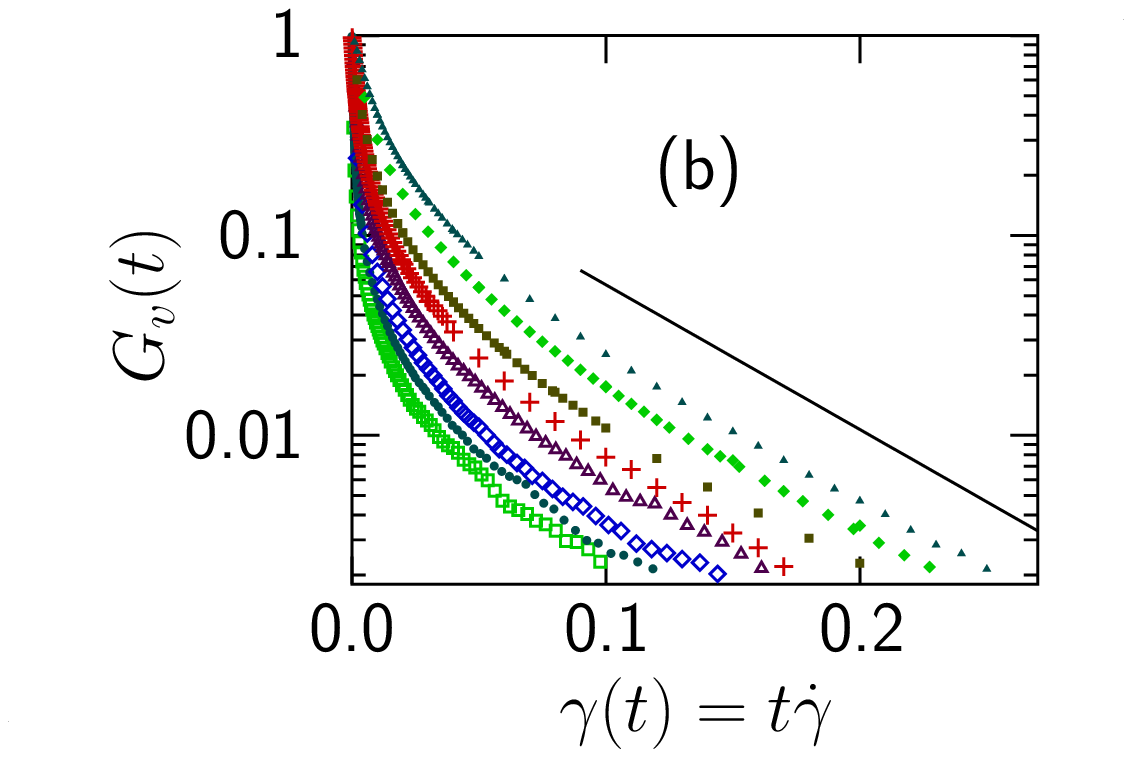} 
  \includegraphics[width=8cm]{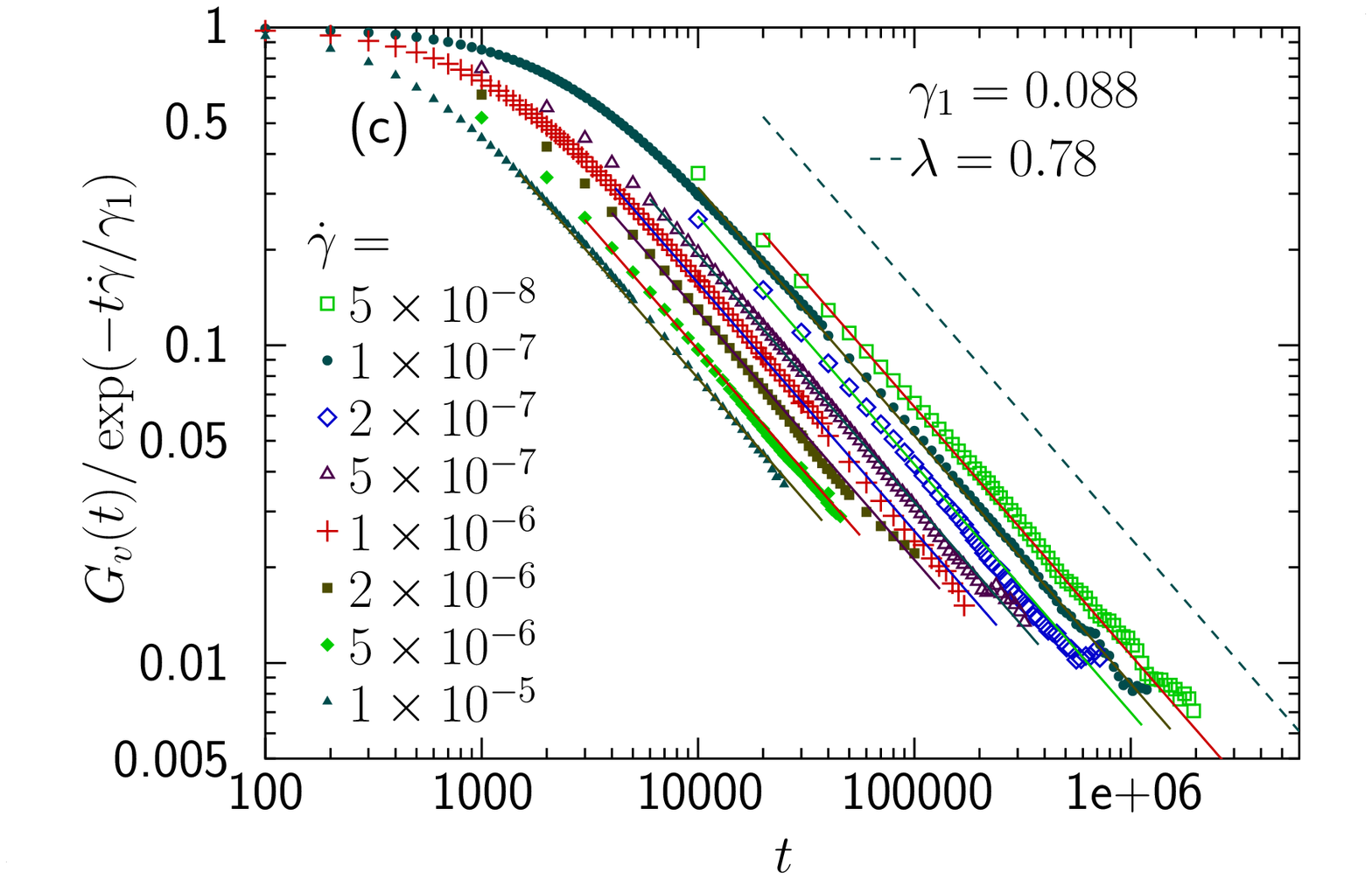} 
  \includegraphics[width=8cm]{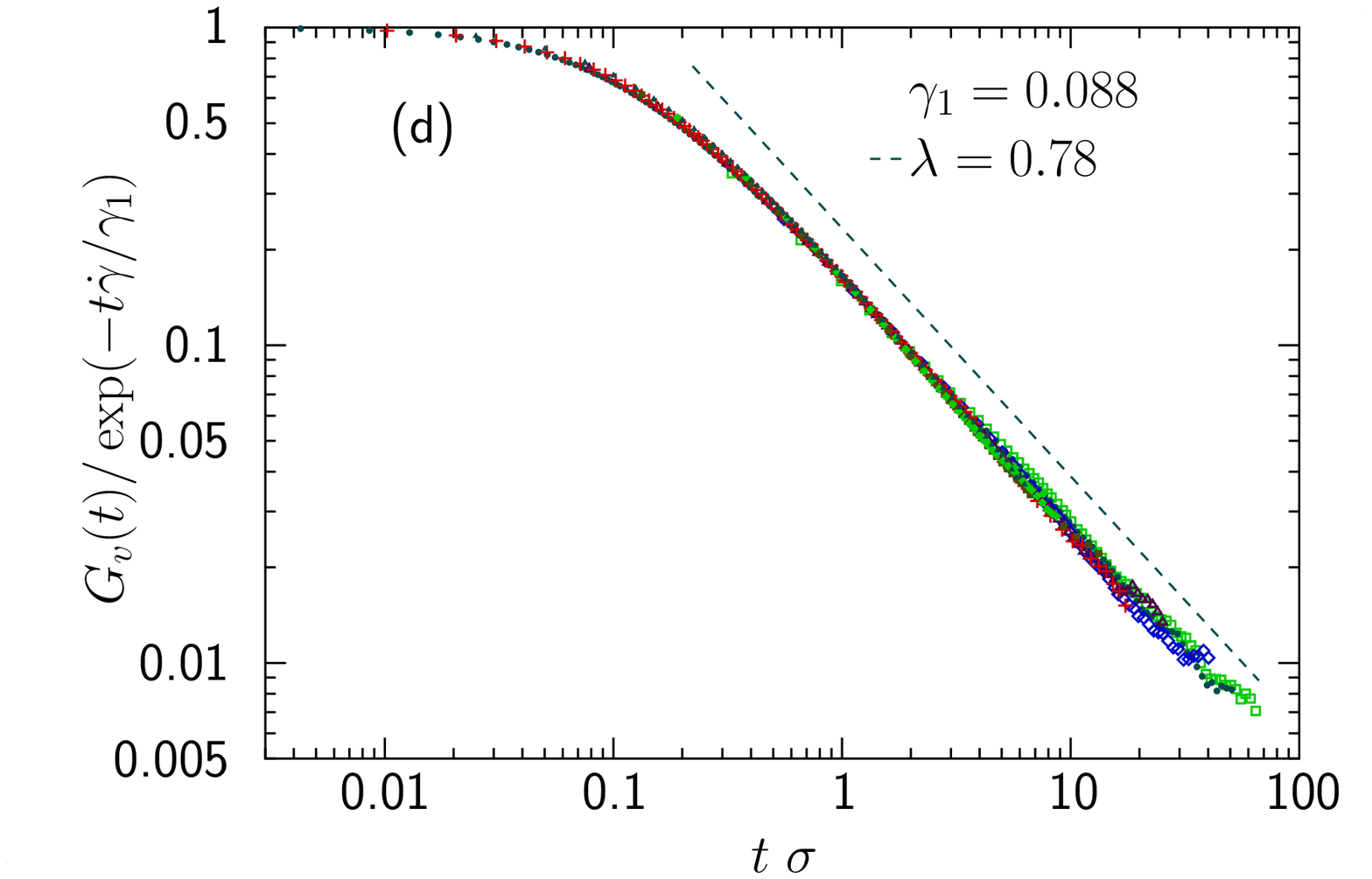} 
  \caption{The velocity auto-correlation function. Panel (a) shows the
    normalized auto-correlation function versus time and panel (b) is the same
    data against strain $\gamma(t)$. The key to our further analysis is the
    rectilinear behavior at large $\gamma$ in panel (b) which suggests an
    exponential decay at large $\gamma$, $\sim e^{-\gamma(t)/\gamma_1}$. The
    line corresponds to $\gamma_1=0.06$. Panel (c) is the same data but now
    compensated for such an exponential decay with $\gamma_1=0.088$ (from
    fitting to \Eq{eq:Gv-scale}).  Panel (d) shows the collapse, now with
    $t\sigma$ as the scaling variable.  The scaling function obeys $\tilde{G}(x)
    \sim x^{-\lambda}$ for large values of its argument.}
  \label{fig:Gv}
\end{figure}

To examine this behavior we turn to the velocity auto-correlation function which
is shown in \Fig{fig:Gv}(a) for a range of shear rates. The same data is shown
also in panel (b), but now plotted against $\gamma(t) = t\gdot$ with a linear
scale on the $x$ axis. From this figure it seems that $\log G_v$ at large
$\gamma$ behaves linearly with similar slopes for different $\gdot$, which
suggests an exponential decay, $\sim e^{-\gamma(t)/\gamma_1}$.  We take this to
suggest that $G_v(t, \gdot)$ may be written
\begin{equation}
  G_v(t, \gdot) = \Gint(t, \gdot)\; e^{-t\gdot/\gamma_1},
  \label{eq:Gint}
\end{equation}
which means that $G_v$ is a product of an exponential decay governed by the
externally imposed time scale $t_1=\gamma_1/\gdot$ and a function $\Gint$, which
captures the internal relaxational dynamics.

In the attempt to make sense of this data we found that a certain choice of
$\gamma_1$ in \Eq{eq:Gint} leads to a great simplification.  It turns out that
$\Gint(t,\gdot)$ is then a function of the combination $t\gdot^\kappa$,
\begin{displaymath}
  \Gint(t,\gdot) = \tilde{G}(t\gdot^\kappa),
\end{displaymath}
and, furthermore, that this function at large values of its argument behaves
algebraically, $\tilde{G}(x) \sim x^{-\lambda}$, see \Fig{fig:Gv}(c).  This
defines a new fundamental exponent that characterizes the relaxation dynamics,
$\sim t^{-\lambda}$, though---for the accessible shear rates---it is to some
extent masked by the exponential decay.  To get unbiased values for these
parameters, we fitted all data in the range $0.002< G_v(t,\gdot) <0.3$ to
\begin{equation}
  G_v(t,\gdot) = A\; (t\gdot^\kappa)^{-\lambda} e^{-t\gdot/\gamma_1},
  \label{eq:Gv-scale}
\end{equation}
with $A$, $\gamma_1$, $\lambda$, and $\kappa$ as free parameters. We find
$\gamma_1=0.088$, $\lambda=0.78$, and $\kappa=0.384$. The solid lines in
\Fig{fig:Gv}(c) show $A(t\gdot^\kappa)^{-\lambda}$ for different $\gdot$.

In this expression, $\gdot^{-\kappa}$ assumes the role of a characteristic time
for the internal relaxation. The observation that ${C}\sigma$ has the dimension
of inverse time together with the good numerical agreement between
$\kappa=0.384$ and $q_\sigma=0.386$ (recall that $\sigma\sim \gdot^{q_\sigma}$)
suggests that $\gdot^{\kappa}$ can be substituted with $\sigma$ such that the
scaling function may be written $\tilde{G}(t\sigma)$.  Panel (d) shows the
collapse when plotting against the scaling variable $t\sigma$. Note that
$\gamma_1$ is the only adjustable parameter in this plot.

\begin{figure}
  \includegraphics[width=8cm]{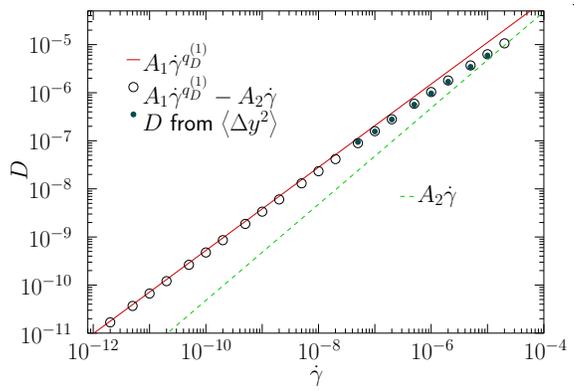} 
  \caption{Behavior of the diffusion constant. In the limit of low shear rate we
    expect the behavior which is given by the solid line, $D = A_1
    \gdot^{q_D^{(1)}}$. The open circles include the corrections to this
    behavior as given by \Eq{eq:qD}. Note the good agreement with the measured
    $D$ from \Fig{fig:D-gdot}. We conclude that $q_D\approx 0.785$ from the
    solid symbols (cf.\ \Fig{fig:D-gdot}), is only an effective exponent that
    describes the behavior in a limited range of shear rates.}
  \label{fig:D-exponent}
\end{figure}

We now like to determine the relation between $q_D$ and the exponents $q_\sigma$
and $\lambda$ that characterize the scaling of $G_v(t,\gdot)$. In the first
approximation we neglect the saturation of $G_v$ at small $t$, and, in effect,
assume that \Eq{eq:Gv-scale} holds down to $t=0$. This gives
\begin{displaymath}
  D \sim \gdot^{1+q_\sigma} \gdot^{-\kappa\lambda}
  \left(\frac{\gdot}{\gamma_1}\right)^{\lambda-1}
  \int_0^\infty dx\; x^{-\lambda} e^{-x}
  \sim \gdot^{q_\sigma+\lambda-\kappa\lambda},
\end{displaymath}
which leads to $q_D^{(1)} \equiv q_\sigma+\lambda-\kappa\lambda = \lambda +
(1-\lambda) q_\sigma = 0.865$. This is the expected behavior as
$\gdot\rightarrow0$, but since it is derived from a simplified $G_v(t,\gdot)$
and the result is well above $q_D=0.785$ from \Fig{fig:D-gdot}, we next try to
take the saturation of $G_v(t,\gdot)$ at small $t$ into account and write
\begin{displaymath}
  G_v(t,\gdot) = \left\{
  \begin{array}{ll}
    1, \quad & t\gdot^\kappa < \xi_0, \\
    A\; (t\gdot^\kappa)^{-\lambda} e^{-t\gdot/\gamma_1}, & t\gdot^\kappa > \xi_0.
  \end{array}
  \right.
\end{displaymath}
Assuming that $e^{-t\gdot/\gamma_1} \approx 1$ at $t=\xi_0/\gdot^\kappa$ (which
holds to a good approximation up to our highest shear rate, $\gdot=10^{-5}$) the
function is continuous at $\xi_0$ if $A=\xi_0^\lambda$.  We then get
\begin{equation}
  \label{eq:qD}
  D = A_1 \gdot^{q_D^{(1)}} - A_2 \gdot,
\end{equation}
with $A_1 = 0.235$ and $A_2 = 0.480$. As shown in \Fig{fig:D-exponent} this
expression (open circles) approaches $D\sim\gdot^{q_D^{(1)}}$ (solid line) at
small $\gdot$ whereas there is an appreciable difference with a smaller slope at
larger $\gdot$. The diffusion constant from \Fig{fig:D-gdot} is shown as solid
dots. Note that it agrees well with the open circles from \Eq{eq:qD}. It
therefore seems that the measured exponent $q_D=0.785$ is not the true
asymptotic behavior.

A central conclusion from our analysis is that the full $G_v(t,\gdot)$ is
approaching an algebraic behavior $\sim t^{-\lambda}$ as $\gdot\rightarrow0$.
We now speculate that the algebraic behavior is related to the finding from
quasistatic simulations that individual plastic events often are avalanches of
elementary flips \cite{Maloney_Lemaitre:2006, Lemaitre_Caroli,
  Goyon_COA_Bocquet:Nature2008, Lerner_Procaccia}. The reason for making this
connection is ideas from self-organized criticality---with the paradigmatic
sandpile model---that a driven system can automatically adjust itself such that
there are avalanches on all length and time scales, which would be seen through
power laws. With a sufficiently low shear rate (say $\gdot=10^{-10}$ or
$10^{-9}$) there would be time for the avalanches to occur one at a time and
evolve according to their own dynamics. At higher shear rates other effects
appear that kill off the avalanches. One possible mechanism is that a new
avalanche interferes with an existing one and thereby destroys its internal
dynamics. Another possibility is that it is simply the shearing of the
simulation box that destroys the correlations.

To conclude, we have found that the velocity auto-correlation function is
governed by two different time scales. With $t_1 = \gamma_1/\gdot$ from the
externally applied shear rate and $t_\mathrm{int} = \sigma^{-1} \sim
\gdot^{-q_\sigma}$ for the internal relaxation, the velocity auto-correlation
function is $G_v(t, \gdot) = \tilde{G}(t/t_\mathrm{int}) e^{-t/t_1}$, where
$\tilde{G}(x) \sim x^{-\lambda}$ for large $x$ and $\lambda$ is a new
fundamental exponent. This also leads to the desired expression for $q_D$ in
terms of two fundamental exponents, $q_D = \lambda + (1-\lambda)q_\sigma$.  We
speculate that this algebraic decay is related to avalanches of elementary
flips, and could be a manifestation of self-organized criticality.

I thank P. Minnhagen and S. Teitel for helpful discussions. This work was
supported by the Swedish Research Council and the High Performance Computer
Center North.


\end{document}